# CVD of $CrO_2$ Thin Films: Influence of the Deposition Parameters on their Structural and Magnetic Properties

A.F. Mota[1], A.J. Silvestre[2], P.M. Sousa[1], O. Conde[1a], M.A. Rosa[3], M. Godinho[3]

[1]Dep. Física/ICEMS, Faculdade de Ciências da Universidade de Lisboa, PT 1749-016 Lisboa

[2]Instituto Superior de Engenharia de Lisboa/ICEMS, PT 1950-062 Lisboa

[1]Dep. Física/CFMC, Faculdade de Ciências da Universidade de Lisboa, PT 1749-016 Lisboa

[a]oconde@fc.ul.pt

**Abstract.** This work reports on the synthesis of $CrO_2$ thin films by atmospheric pressure CVD using chromium trioxide ($CrO_3$) and oxygen. Highly oriented (100) $CrO_2$ films containing highly oriented (0001) $Cr_2O_3$ were grown onto $Al_2O_3$(0001) substrates. Films display a sharp magnetic transition at 375 K and a saturation magnetization of 1.92 $\mu_B$/f.u., close to the bulk value of 2 $\mu_B$/f.u. for the $CrO_2$.

**Introduction**

Magnetic materials exhibiting a high degree of spin polarization are being actively investigated for their potential use in spintronic devices. Among them, chromium dioxide ($CrO_2$) is very attractive because it is a half-metal fully spin polarized at the Fermi level with Curie temperature above room temperature ($T_c$=393 K) [1-4]. $CrO_2$ has the rutile structure with a tetragonal unit cell (a=b=0.4421 nm, c=0.2916 nm) consisting of two formula units, the Cr ions being in the $Cr^{4+}$ oxidation state, and a magnetic moment of 2 $\mu_B$/f.u. [5].

The ability to produce $CrO_2$ films is crucial both for applications and fundamental studies. Nevertheless, it remains a challenging task to grow thin films of $CrO_2$ due to its metastability. $CrO_2$ easily decomposes into the stable $Cr_2O_3$ insulating antiferromagnetic oxide phase. Up to now, chemical vapour deposition (CVD) based on the methodology described by Ishibashi et al. [6] seems to be the most successful technique to produce such films. High quality $CrO_2$ films have been deposited on single-crystal rutile ($TiO_2$) [6-10], which has the same crystal structure and nearly the same lattice parameters as $CrO_2$. Sapphire has also been used as substrate material [11], however the films grown on this substrate have been shown to be contaminated with $Cr_2O_3$.

In this work, we focus on the synthesis and properties of $CrO_2$ thin films on $Al_2O_3$(0001) substrates by atmospheric pressure CVD. The study of the deposition process as a function of the deposition time allowed us to give an explanation for the occurrence of a $Cr_2O_3$ layer at the $CrO_2$ film/substrate interface on the basis of structural effects.

**Experimental**

The $CrO_2$ films were deposited onto $Al_2O_3$(0001) substrates by atmospheric pressure CVD following the method of Ishibashi et al. [6]. However, instead of the two-zone furnace used by these authors, we used a quartz tube placed inside a single-zone furnace and an independent control of the substrate temperature, similar to the system described in ref. [12]. Chromium trioxide ($CrO_3$) powder (purity 99.9%) was used as precursor and loaded into a stainless steel boat aligned with the substrate holder, 6 cm apart from it. Prior to their insertion into the reactor, the substrates were ultrasonically cleaned in organic solvents, rinsed in distilled water and dried with a $N_2$ flux. Oxygen (purity 99.999%) was used as carrier gas.

Films were grown for the experimental parameters presented in table 1. In order to avoid the deposition of any kind of impurities during the initial stages of the deposition process, the substrate was always heated up to the deposition temperature before the melting of the $CrO_3$ precursor occurred ($T_m$=196 °C). Best results were obtained for substrate temperature $T_s$ = 390 °C, precursor temperature $T_p$ = 260 °C and oxygen flow rate $\phi_{O2}$ = 280 sccm. In this paper we will concentrate on the analysis of films grown with these experimental parameters for deposition times ranging from 1 to 6 h.

**Table 1** – CVD experimental conditions used to deposit $CrO_2$ thin films

| *Experimental parameters* | *Values* |
|---|---|
| Substrate temperature, $T_s$ [°C] | 383 - 400 |
| Precursor temperature, $T_p$ [°C] | 260 - 295 |
| Oxygen flow rate, $\phi_{O2}$ [sccm] | 180 - 500 |
| Deposition time, $t_{dep}$ [h] | 1 - 6 |

The morphology and microstructure of the films were analysed by scanning electron microscopy (SEM). Their structure was studied by X-ray diffraction (XRD) with Cu K$\alpha$ radiation and by micro-Raman spectroscopy using the 514.5 nm excitation line of an $Ar^+$ laser. The magnetic characterization was carried out using a commercial SQUID magnetometer. Field cooling curves (H=50 Oe) for increasing temperatures between 35-400 K and hysteresis loops both for parallel and perpendicular applied magnetic fields at 35 K, were obtained for representative samples.

**Results and discussion**

**Surface morphology and microstructure.** The deposited films present good adherence, are homogeneous and exhibit a black shiny metallic colour. SEM analyses of different films have shown that their morphology consist of a uniform and very fine grain structure only resolved at high magnification (Fig. 1). For films grown with $T_s$=390 °C, $T_p$=260 °C, $\phi_{O2}$=280 sccm and $t_{dep}$=6 h a uniform thickness of 260±10 nm was measured, yielding a deposition rate of 0.72±0.03 nm.min$^{-1}$.

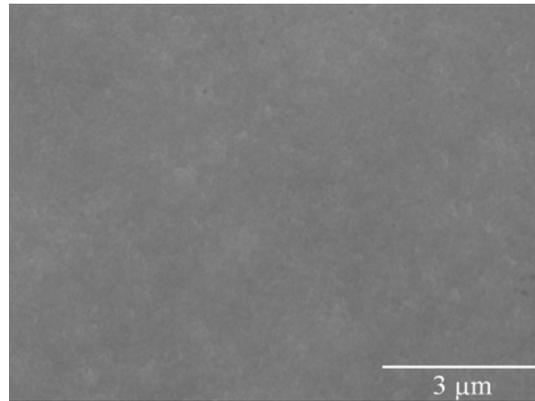

**Fig. 1** – Scanning electron micrograph of a film deposited with $T_s$=390 °C, $T_p$=260 °C, $\phi_{O2}$=280 sccm and $t_{dep}$=6 h. Magnification: ×8000

**Structural analysis.** Fig. 2 displays the XRD pattern of a film grown during 6 hours. Concerning $CrO_2$ deposition, it can be seen that only the (200) and (400) diffraction peaks of $CrO_2$ are clearly identified, revealing highly a-axis oriented film growth. The diffraction pattern also shows the presence of peaks from the $Cr_2O_3$ phase, (006) and (0012), although less intense than those of $CrO_2$, indicating that the chromia phase is also highly oriented.

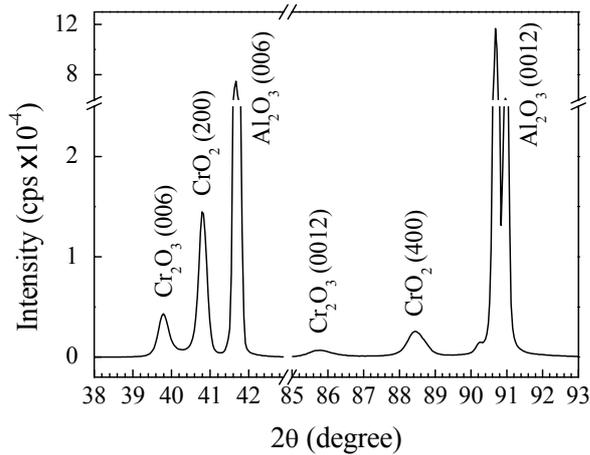
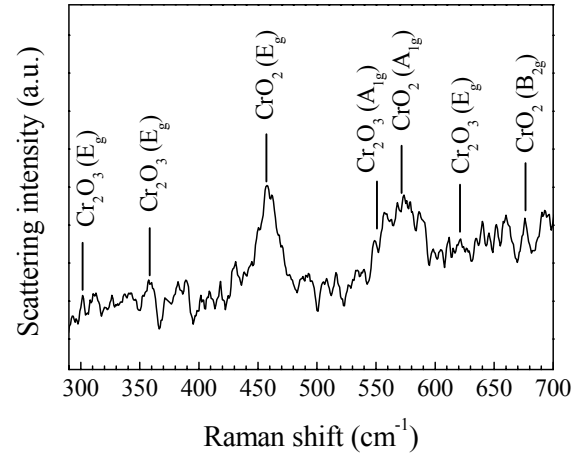

**Fig. 2** – X-ray diffractogram of a $CrO_2$ film deposited with $T_s$=390 ºC, $T_p$=260 ºC, $\phi_{O2}$=280 sccm and $t_{dep}$=6 h.

**Fig. 3** – Micro-Raman spectrum of the film shown in Fig. 2.

In order to understand the location of the $Cr_2O_3$ phase in the sample, Raman microprobe studies were carried out by recording Raman spectra over different zones on the film surface. Fig. 3 shows a typical micro-Raman spectrum where two characteristic bands of the chromium dioxide are clearly observed at 458 cm$^{-1}$ ($E_g$ mode) and 573 cm$^{-1}$ ($A_{1g}$ mode) Raman shifts [13].

If some $Cr_2O_3$ were on the film surface it would be easily detected at the Raman shifts labeled on the figure, as the Raman scattering cross-section of $Cr_2O_3$ is much larger than that of $CrO_2$. Since this is not the case for the spectrum in Fig. 3, we can speculate that the $Cr_2O_3$ forms a layer at the $CrO_2$ film/substrate interface. A set of experiments was carried out varying the deposition time, in order to study the evolution of both chromium oxide phases as the films grow.

Fig. 4 presents X-ray diffractograms of films grown for different deposition times. After one hour of deposition only the $Cr_2O_3$ phase is observed via the (006) and (0012) diffraction lines. As deposition time increases, the intensities of the (200) and (400) peaks of the $CrO_2$ phase continuously increase in relation to those of $Cr_2O_3$ that stabilise after 2 hours. This result is consistent with $Cr_2O_3$ being formed as an interface layer between $CrO_2$ and the sapphire substrate.

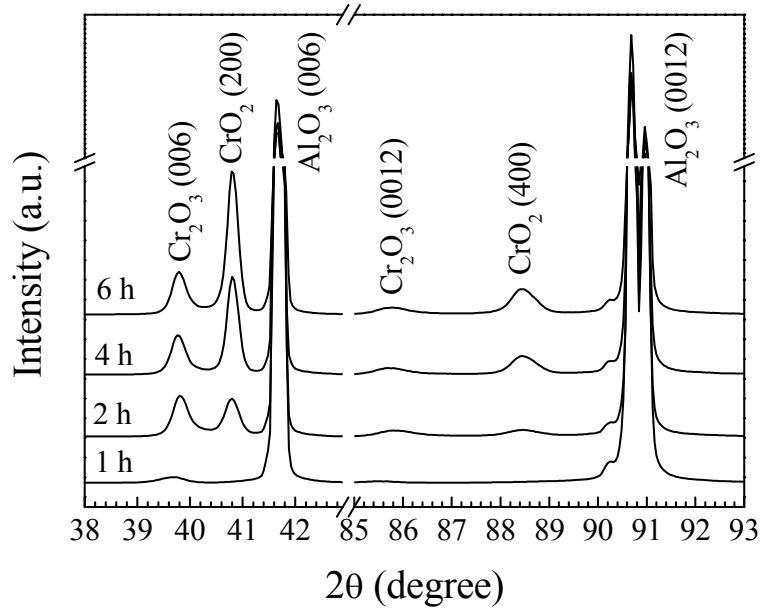

**Fig. 4** – X-ray diffractograms of films deposited with $T_s$=390 ºC, $T_p$=260 ºC, $\phi_{O2}$=280 sccm and deposition times ranging from 1 to 6 h.

The preferred growth of $Cr_2O_3$ oxide phase in the early stages of the deposition process was previously reported by M. Rabe et al. [11]. These authors considered that $Cr_2O_3$, which is the stable phase in the Cr-O phase diagram at atmospheric pressure, develops between the onset of the precursor temperature at ~260ºC and the instant at which the substrate attains the required temperature of ~390 ºC for film growth. By contrast, in our experiments the deposition temperature is always attained before the $CrO_3$ melts at ~196 ºC. Therefore, the growth of the $Cr_2O_3$ layer onto the sapphire substrate cannot be attributed to inadequate temperature differences between the precursor and the growing film. Instead, it can be explained by the close structural match between $Cr_2O_3$ and the $Al_2O_3$ substrate. Both compounds have hexagonal structure and close interatomic distances in the (001) plane: $a_{Cr2O3}$ = 0.4959 nm and $a_{Al2O3}$ = 0.4759 nm, leading to a lattice mismatch of only 4.2%. Therefore, the chromia layer is slightly compressed in the film plane and expanded along the c-axis: the c parameter calculated from the diffractogram of the film grown during the 1st hour (Fig. 4) is 0.15% larger than the expected bulk value.

However, this small mismatch value is responsible by an increase of the layer energy as elastic strain energy. As the film thickness increases, the accumulated energy increases up to a critical thickness value for which the excess energy is sufficient to induce transition to a new phase, i.e. $CrO_2$. It is interesting to note that the distance between the atomic positions with coordinates $10\bar{1}0$ and $\bar{1}\bar{1}20$ (0.8589 nm) differs of only 1.8% from the triple of the c parameter of $CrO_2$ lattice. This may justify an orientation relationship of the type $(100)_{CrO2} \parallel (001)_{Cr2O3}$ and $[001]_{CrO2} \parallel [210]_{Cr2O3}$, thus explaining the XRD results. Therefore, the growth of a $Cr_2O_3$ buffer layer between the $CrO_2$ and the sapphire substrate seems to be an intrinsic feature of the chosen film/substrate system.

**Magnetic characterisation.** Despite the presence of the $Cr_2O_3$ buffer layer, the $CrO_2$ films display a strong ferromagnetic behaviour. Fig. 5 shows the magnetization curve as a function of temperature for the film deposited during 6 h. As can be seen, a sharp magnetic transition is observed for T ≈ 375 K as determined from the inflection point of the magnetization versus temperature curve[a]. Fig. 6 shows the magnetic hysteresis loop of the same film, measured at 35 K with the applied magnetic field parallel to the substrate surface. From this curve a coercive field of 110 Oe and a saturation magnetization of 624 emu.cm$^{-3}$ were determined, the later corresponding to 1.92 $\mu_B$/f.u. close to the bulk value of 2 $\mu_B$/f.u. for the $CrO_2$. The analysis of hysteresis loops recorded both for parallel and perpendicular applied magnetic fields showed that the magnetization easy-axis is in the film plane.

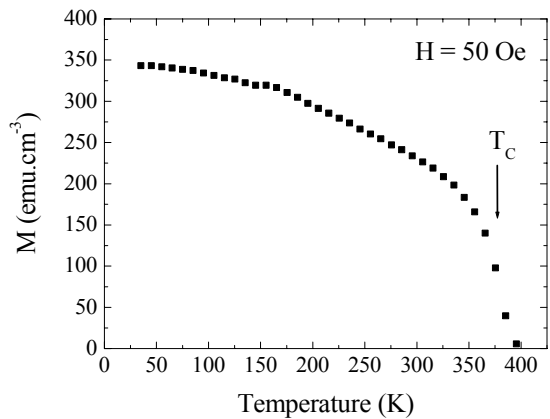
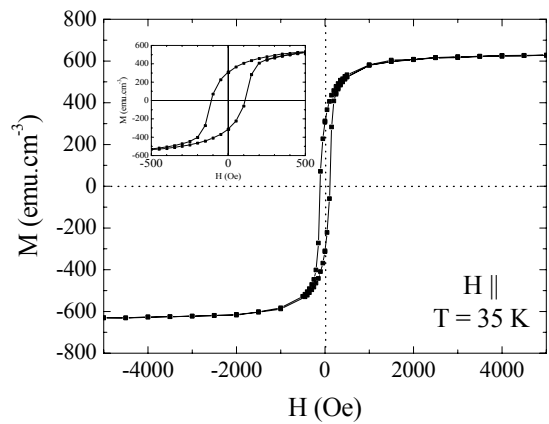

**Fig. 5** – Magnetization vs. temperature for a $CrO_2$ film deposited with $T_s$=390 ºC, $T_p$=260 ºC, $\phi_{O2}$=280 sccm and $t_{dep}$=6 h, measured at H=50 Oe.

**Fig. 6** – Magnetic hysteresis curve of a $CrO_2$ film deposited with $T_s$=390 ºC, $T_p$=260 ºC, $\phi_{O2}$=280 sccm and $t_{dep}$=6 h, measured at T=35 K with the field parallel to the substrate surface. The inset shows the low field hysteresis loop for the same film.

**Conclusions**

Highly oriented $CrO_2$ thin films were deposited onto $Al_2O_3$(0001) substrates by atmospheric pressure CVD, using $CrO_3$ and $O_2$. The growth of $Cr_2O_3$ oxide as a buffer layer at the $CrO_2$/sapphire interface was shown and explained on the basis of structural match between the oxide and the substrate. Despite the existence of a $Cr_2O_3$ interlayer, the films display a sharp magnetic transition at 375 K and a saturation magnetization of 1.92 $\mu_B$/f.u..

---

[a] If $T_c$ is taken as the temperature for which M→0 [10], the value deduced from Fig. 5 is ~396 K.

**Acknowledgements**
This work is supported by FCT (Portugal) contract POCTI/CTM/41413/2001 and EU contract FENIKS: G5RD-CT-2001-00535.